\documentclass{elsart}
\usepackage{epsfig}
\begin{document}
\newcommand{\cal}{\it}
\begin{frontmatter}
\title{Polynomial Lie algebra methods in solving the
second-harmonic generation model: some exact and approximate
calculations\thanksref{ths}}
\thanks[ths]{A financial support of the work from
the Russian Foundation for Basic Research under grant
No.00-02-81023 Bel 2000\_a is acknowledged.}

 \author[fian,jinr]{V.P. Karassiov}
 \author[jinr]{A.A. Gusev}
 \author[jinr]{S.I. Vinitsky}
 \address[fian]{P.N. Lebedev Physical Institute, Leninsky Pr. 53, Moscow,
 117924 Russia, email: vkaras@sci.lebedev.ru  (contact address)}
 \address[jinr]{JINR, Dubna, Moscow region, Russia,\\
email: vinitsky@thsun1.jinr.ru}

\begin{abstract}
We compare exact and $SU(2)$-cluster approximate calculation
schemes to determine dynamics of the second-harmonic generation
model using its reformulation in terms of a polynomial Lie
algebra $su_{pd}(2)$ and related spectral representations of the
model evolution operator realized in algorithmic forms. It
enabled us to implement computer experiments exhibiting a
satisfactory accuracy of the cluster approximations in a large
range of characteristic model parameters.
\end{abstract}
\end{frontmatter}
\section{ Introduction}
 During the last half-century a great attention is being paid to studies
 of different quantum models of nonlinear optics
 since they enable to  reveal new physical effects and phenomena (see,
 e.g.,  \cite{1,2,3,4,5,6,7,8,9,10,11,12,13}  and references therein). In view of the Hamiltonian
 nonlinearity these models are mainly analyzed with the help of the numerical
 calculations \cite{9,19,20}
 or some linearization procedures \cite{4,7,8} which are not adapted to reveal
 many peculiarities of model dynamics \cite{2,5,9,11,12,13}. However, recently  a new
 universal Lie-algebraic approach, essentially improving both analytical
 and numerical solutions of physical problems, has been suggested in \cite{14}
 and developed in \cite{6,11,12,15} for the class of nonlinear quantum models
 whose Hamiltonians $H$ have invariance groups $G_i:\, [G_i, H]=0$. This
 approach is based on reformulations of models under study in terms of
 dual algebraic pairs $(G_i,g^D)$  \cite{12,14} where dynamic symmetry algebras
 $g^D$ are  polynomial Lie algebras (PLA) $g_{pd}: g^{D}=g_{pd}$
 completely  describing model dynamics. Specifically, for a wide class
 of quantum-optical models  with $g^{D} = g_{pd}=su_{pd}(2)$  both exact
 and approximate methods were developed in \cite{6,11,12,14,15} to obtain adequate
 (for physical applications) forms of the model  evolution operators $U_H(t)$.
 Exact methods include appropriate Lie-algebraic path-integral schemes
 \cite{12} and $su_{pd} (2)$-techniques \cite{6,14,15} based on solving difference and
 difference-differential equations which  are  fairly complicated  for
 practical calculations because they yield only algorithms rather
 than closed analytical expressions  for solving physical problems.
 Therefore, for this aim two analytical
 approximations of $U_H(t)$ were found in \cite{11}. They enable to
 examine model dynamics at cluster  quasiclassical levels described by
 "large" values of certain characteristic parameters \cite{11,12}. However,
 up to now systematic examinations of their accuracy and
 efficiency were absent.

 In the present work we cancel in part this lacuna for the simplest
 and most widely examined (see, e.g. \cite{1,3,4,6,7,8,16,17,18,19,20} and references
 therein) example of such models describing process of second-harmonic
 generation  which is given  by the  Hamiltonian
%=====
\begin{equation} H = \hbar\left[ \omega_0 a_0^+ a_0 +
\omega_1 a_1^+ a_1  + g {a^+_{1}}^2 a_0 + g a_{1}^2 a^+_0\right],
\label{1.1a}\end{equation}
%=============
and by the  Fock space $ L_F(2)=Span\{|\{n_1,n_0\}\rangle\propto
(a_1^+)^{n_1}(a_0^+)^{n_0}|0\rangle\}$
% \label{1.1b)$$
as the model Hilbert space $L(H)$; here  $a_{i}, a _{i}^+$ are
operators of field modes with frequencies $\omega_i$, $g$ are
coupling constants. Classical solutions of the model (\ref{1.1a}) were
first found (in terms of elliptical functions) in \cite{1},
and later its different quantum
features were examined by many investigators (see, e.g.,
\cite{3,4,6,7,8,11,12,13,14,15,16,17,18,19,20} and references therein). Specifically, in
\cite{11,17,19} some peculiarities, like the collapse-revival
phenomena in the Janes-Cummings model \cite{2}, were revealed for
its model dynamics at the quantum and quasiclassical levels.

The work is organized as follows.  After some preliminaries
(Section 2) we give exact and approximate calculation schemes and
routines to determine "diagonal" (spectral) representations of
the model evolution operator by means of the
$su_{pd}(2)$-techniques \cite{14,15} (Section 3). In Section 4 we
implement computer experiments to exhibit their efficiency and to
investigate the accuracy of appropriate quasiclassical
approximations obtained in \cite{11}. In Conclusion  we briefly
discuss some of ways to improve the results obtained.
% 2   %%%%%%%%%%%%%%%%%%%%%%%%%%%%%%%%%%%%%%%%%%%%%%%%%%%%%
\section{The $su_{pd}(2)$ formulation of the  model
 and a related "diagonal" representation of its dynamics}
%%%%%%%%%%%%%%%%%%%%%%%%%%%%%%%%%%%%%%%%%%%%%%%%%%%%%

 As was shown in \cite{12,14}, the invariance  of the model
Hamiltonian (\ref{1.1a})
with respect to the group $G_i=C_2\times \exp(i\lambda R_1), C_2=
\{\exp(i r N_1): r=0,1\}$ enables to re-formulate the model under
study
 in terms of two sets of collective operators:
 integrals of motion $R_1= ( N_1+2 N_0)/ 3,\, K= N_1- 2[N_1/2]\,(N_{i}=
 a_{i}^+a_{i}, [A]$ is the entire part of $A$)
 and $G_i$-invariant dynamic variables $V_{\alpha=0,\pm}$:
  \begin{eqnarray} V_+ =  a^{+2}_{1} a_0, \quad V_- = a_{0}^+a_{1}^2,
 \quad V_{0} =  (N_1 - N_0) /3,
  \label{2.1}\end{eqnarray}
They obey the commutation relations
%%%%%%%%%
 \begin{eqnarray}&&[V_0, V_{\pm}]= \pm V_{\pm},\quad
 [V_{\alpha}, R_1] = 0 = [V_{\alpha}, K],  \nonumber \\\;&& [V_-, V_+] = \phi (V_0; R_1)
 \equiv \psi (V_0+1; R_1) -\psi (V_0; R_1), \nonumber \\&&
 \psi (V_{0};R_1)
 =(R_1 + 2 V_{0})(R_1 + 2 V_{0}-1) (R_1 + 1 -  V_{0}),
 \label{2.2a}\end{eqnarray}
 that identifies $V_{\alpha}$ as generators of PLA $su_{pd}(2)$ having
 the Casimir operator
  \begin{eqnarray}&&\Psi (R_0=(K-R_1)/2)\equiv \psi (V_0;R_1)-V_+ V_- = \psi (V_0+1;R_1)-V_-
 V_+ , \nonumber \\ {}&& [V_{\alpha},\Psi (R_0)]=0, \label{2.2b}\end{eqnarray}
 acting on $L(H)$ complementarily to $G_i$  (in view of the relationship
 $K = 2 R_0 + R_1$) and, hence, forming dynamic symmetry algebra $g^D$ in
 the dual algebraic pair $(G_i,g^D)$ \cite{12,14}.

In terms of these collective operators
 the Hamiltonian (\ref{1.1a})  is expressed in the form \cite{14}
%%%%%%%%%%%%%%
\begin{eqnarray} &&H = \hbar [\Delta V_0 + g V_+ + g^* V_- + C(R_1)],\;\nonumber \\ && C = (\omega_1+
\omega_0) R_1,\quad \Delta= 2 \omega_1- \omega_0,\quad
[V_{\alpha}, C] = 0,   \label{2.3a}\end{eqnarray}
%%%%%
and  the Hilbert space $L(H)$ is  decomposed into the infinite
direct sum
 \begin{eqnarray}&&L(H) =  \sum_{k=0,1} \sum_{s=0}^\infty L (k,s),\nonumber \\
  &&L (k,s) =
 \mbox{Span} \{|k,s;f\rangle:\, R_{i=0,1}\, |k,s;f\rangle = l_i
 \,|k,s;f\rangle\},
 \label{2.4a}\end{eqnarray}
%%%%%%%%%%%%%%%%%%
of $G_i\otimes su_{pd}(2)$-irreducible $s+1$ - dimensional
subspaces $L(k,s)$ specified by eigenvalues $l_{i}$ of the
$G_i\otimes su_{pd}(2)$ invariant operators $R_{i} $. Herewith
$l_{i}$ are expressed through the numbers $s, k:\, l_0=({k -
s})/{3},l_1= ({k + 2s})/{3}$ determining, respectively, a maximal
population of the fundamental (pump) mode and a minimal
population of the harmonic within a fixed "optical atom" $L(k,s)$
as it  follows from the structure of the new (collective) basis
in $L(H)$ \cite{11}:
\begin{eqnarray}&&|k,s;f\rangle = N(f;k,s) V_+^f |k,s \rangle = |\{n_1= k + 2f, n_0 =
s-f\}\rangle,\,\nonumber \\&& N(f;k,s) = [{s!
k!}/{(k+2f)!(s-f)!}]^{1/2}, \nonumber \\&& V_0 |k,s;f\rangle =
(l_0+f) |k,s;f\rangle,\nonumber \\&& |k,s\rangle= |n_1= k, n_0=
s\rangle, \quad V_-\,|k,s\rangle = 0,\quad k=0,1,\,\,s\geq 0.
\label{2.4b}\end{eqnarray} Evidently,  Eqs. (\ref{2.4b})
explicitly manifest the $su_{pd}(2)$ cluster structure of the
Fock states $|n_1,n_0\rangle$ and specify $R_0$ as the "lowest"
weight operator and $|k,s \rangle$ as the "lowest" weight state.

 This $su_{pd}(2)$ cluster reformulation of the model enables one  to
 use the $su_{pd}(2)$ formalism for getting representations of the model
 evolution operator $U_H(t)=\exp(- iHt/\hbar )$ which facilitate
 analysis of the model dynamics and calculations of temporal dependences
 $\langle O(t)\rangle=\mbox{Tr} [U_{H}(t) \rho U^{\dagger}_{H}(t) O]$
for arbitrary physical observables $O$; $\rho=\rho(0)$ is the
density
 operator of the initial quantum  state \cite{6,11}.
 One of such adequate representations has a diagonal form \cite{11,12}
 \begin{eqnarray}U_H(t) = \sum_{k,s;v} \,e ^{-it\epsilon (k,s;v)}
 |E_v(k,s) \rangle\,\langle E_v(k,s)|, \nonumber \\ E_v(k,s)= \hbar
 \epsilon (k,s;v)= \hbar [C(l_1) + \lambda_v(k,s)], \label{2.5}\end{eqnarray}
 %%%
 where $E_v(k,s)$ and $\{|E_v(k,s)\rangle\}$ are  eigenvalues and complete
sets of orthonormalized eigenvectors of the Hamiltonian (\ref{2.3a})
respectively:
 %%%
 \begin{equation}H |E_v(k,s)\rangle = E_v(k,s)\, |E_v(k,s)\rangle,
 \quad \!\!
 \langle E_v(k,s)|E_{v'}(k',s')\rangle =\delta_{k,k'} \delta_{s,s'}
 \delta_{vv'}. \label{2.6}\end{equation}
 %%%
In the  basis $\{|E_v(k,s)\rangle\}$  quantum expectations
$\langle O(t)\rangle$ are given  as follows \cite{11}:
\begin{eqnarray}\langle O(t)\rangle= \sum_{k,s,v; k',s' v'} &\langle E_v(k,s)|\rho
|E_{v'}(k',s')\rangle \langle E_{v'}(k',s')| O
|E_{v}(k,s)\rangle\,\times\nonumber \\ & \times e^{i t [ \epsilon
(k',s';v') -\epsilon (k,s;v) ]}\,,    \label{2.7}\end{eqnarray}
 and, hence, the main problem is in solving the eigenproblem (\ref{2.6}).

% 3   %%%%%%%%%%%%%%%%%%%%%%%%%%%%%%%%%%%%%%%%%%%%%%%%%%%%%
 \section{ Lie-algebraic schemes for finding diagonal $su_{pd}(2)$ forms
  of the model evolution operator: an exact algorithm  and
 cluster quasiclassical approximations}
In the case, when the PLA structure polynomial $\psi (V_0;R_1)$
is quadratic in $V_0$  and PLA $su_{pd}(2)$ is reduced to the
familiar Lie algebra $su(2)= Span\{Y_0,Y_\pm:\,[Y_0,Y_\pm]=\pm
Y_\pm,\,[Y_-,Y_+]=2 Y_0 \}$ the eigenvalue problem (\ref{2.6}) is
solved exactly with the help of the $SU(2)$ displacement
operators $ S_{Y}(\xi=r{g}/{|g|})= \exp(\xi Y_+ - \xi^* Y_-)$
\cite{21} in terms of simple analytical expressions \cite{12}. However, it
is not the case for the model under study in view of the absence
of explicit expressions for matrix elements $\langle k,s; f|
\exp(\sum_i a_i V_i) |k,s; f'\rangle$ \cite{6,15}. Nevertheless, the
$su_{pd}(2)$ formalism enables one get convenient (for physical
applications) calculation schemes, algorithms and analytical
expressions for exact  and approximate solutions of this problem.

A Lie-algebraic scheme for finding exact solutions of the
eigenproblem (\ref{2.6}) is based on looking for eigenfunctions
$|E_v(k,s)\rangle$ on each subspace $L(k,s)$ in the form \cite{14,15}
%%%%
\begin{eqnarray}&&|E_v(k,s)\rangle = \sum_{f=0}^s Q_f^v(k,s)
 |k,s;f\rangle
 =\sum_{f=0}^s \,\tilde Q_f^v(k,s)\, V_{+}^f |k,s\rangle, \;
 \nonumber \\ &&\tilde Q_f^v(k,s)= N(f;k,s)\, Q_f^v(k,s),  \label{3.1a}
\end{eqnarray}
%%%
where amplitudes $Q_f^v(k,s)$ satisfy the orthonormalization and
completeness conditions:
\begin{equation} \sum_{f=0}^s Q_f^v(k,s)\; Q_f^{v'}(k,s) \,=\, \delta_{vv'},
\qquad \sum_{v=0}^s Q_f^v(k,s)\;Q_{f'}^v(k,s)\, =\, \delta_{ff'}.
\label{3.1b} \end{equation}
Then, inserting Eq. (\ref{3.1a}) for $|E_v(k,s)\rangle$ and Eq.
(\ref{2.3a}) for $H$ in Eq. (\ref{2.6}) and using Eqs.
(\ref{2.2a}), (\ref{2.4b}), one gets a set of recurrence
relations at fixed $k=0,1; s=0,1,\ldots$
\begin{eqnarray}&& g^{*}\psi (l_0+f+1; l_1)\tilde Q_{f+1}^{v}(k,s) =
\nonumber\\&&\quad=[\lambda_v(k,s)- \Delta (f + l_0)]\tilde
Q_f^{v}(k,s)) - g\tilde Q_{f-1}^{v}(k,s),\nonumber\\ &&
 f,v=0,\ldots,s,\; \nonumber \\ &&
\psi (l_0+f+1; l_1)= (k + 2f+2)(k + 1  + 2f)(s - f) .    \label{3.2}\end{eqnarray}
These relations along with the boundary conditions $\tilde
Q_{-1}^{v}(k,s) = 0 = \tilde Q_{s+1}^{v}(k,s)$ determine
amplitudes $Q_f^{v}(k,s)$ and eigenenergies $E_{v}(k,s)$ from
solutions of the Sturm-Liouville spectral problem \cite{6,14}
\begin{eqnarray}&&P_{f+1}(\lambda) = [\lambda - \Delta (f + l_0)] P_{f}(\lambda) -
|g|^{2} \psi (l_0+f; l_1) P_{f-1}(\lambda), \label{3.3a}\\ && \,
f=0,\ldots,s; \qquad P_{0}(\lambda) =1, \nonumber\\ &&
 P_{-1}(\lambda) = 0 = [\lambda - \Delta (s +
l_0)] P_{s}(\lambda) - |g|^{2} \psi (l_0+s; l_1) P_{s-1}(\lambda),
\label{3.3b}\end{eqnarray}
for finding non-classical orthogonal (in view of (\ref{3.1b}))
polynomials
\begin{equation}P_{f}(\lambda) =
\frac{(g^* )^f \tilde Q_{f}(k,s;\lambda)}{N^{2}(f;k,s)
\tilde Q_{0}(k,s;\lambda)}=
\frac{(g^* )^f Q_{f}(k,s;\lambda)}{N(f;k,s)Q_{0}(k,s;\lambda)}
  \label{3.4}\end{equation}
of the discrete variable $\lambda$ on the non-uniform lattice
$\{\lambda_v(k,s)\}_{v=0}^s$ \cite{14}. Indeed, Eqs. (\ref{3.3a}), (\ref{3.3b}),(\ref{3.4})
provide the following algorithm for solving the eigenproblem (\ref{2.6}).\\
i) Using the recursive formula (\ref{3.3a}) with the boundary values
from Eq.
(\ref{3.3b}) one calculates the polynomial sequence $\{P_{f}(\lambda)\}_{f=0}^s$.\\
ii) Inserting $P_{s-1}(\lambda), P_{s}(\lambda)$ in the last
equality in (\ref{3.3b}) one gets the algebraic equation with respect
to $\lambda$; its solution yields the  sequence
$\{\lambda_v(k,s)\equiv \lambda_v\}_{v=0}^s$ of admissible values
of the spectral parameter $\lambda$ and the appropriate energy
spectrum   $\{E_v \equiv E_{v}(k,s)\}_{v=0}^s$.\\
iii)  For each value $\lambda_{v}(k,s)$ using
$\{P_{f}(\lambda)\}_{f=0}^s$
 and Eq. (\ref{3.4}) one finds the sequence $\{Q_{f}^{v}(k,s)\equiv Q_{f}(k,s;
 \lambda_v)\}_{f=1}^s$ of all amplitudes as functions of the only
 undetermined quantity $Q_0(\lambda_v)\equiv Q_{0}(k,s; \lambda_v)$
 which, in turn, is found from the normalization condition of Eqs. (\ref{3.1b}).\\
 This algorithm has been realized in \cite{22} with the help of the
 REDUCE procedure SOLVE for $s\leq 100$ and the conventional
 FORTRAN subroutines of EISPARK package (with applying
 the  multiprecision package \cite{23}) for  $s\geq 100$.

 The routine package developed enables us to implement
 numerical calculations of model dynamics for $s\leq 160$ but it is
 unsuitable for practical calculations with larger $s$ in view of
 multiprecision computer limitations. Therefore in \cite{11} an approximate
 analytical solution of the problem (\ref{2.6}) has been suggested. It is given
 by the $SU(2)$-quasiclassical eigenfunctions
 \begin{eqnarray}&&|E^{qc}_v(k,s;\xi=r{g}/{|g|})\rangle = \exp(\xi Y_+ -\xi^*
 Y_-) |k,s;v\rangle \!=\! \sum_f S^j_{f v}(\xi)|k,s; f\rangle,
 \label{3.5a}\\ &&
 Y_0 = V_0-l_0- j,\quad\!\!\!  Y_+= V_+ [2(s + 2Y_0 +2 k +1)]^{-\frac{1}{2}}=(Y_-)^+,
  \quad\!\!\! 2j = s,\nonumber \\ &&Y_{\pm, 0}\, \in\, su (2), \quad
 S^j_{f v}(\xi)\equiv Q_f^{v;\, ap}(k,s)\,=\,(\frac{g}{|g|})^{f-v}\,
 d^{j}_{-j +f,-j +v}(2r),   \label{3.5b}\end{eqnarray}
%%%%%%%%
and eigenenergies
\begin{eqnarray}&& E^{qc}_v(k,s;\xi)
\equiv\langle E^{qc}_v(k,s;\xi)| H |E^{qc}_v(k,s; \xi)
\rangle=\hbar [C(l_1)+ \lambda_v^{qc}(k,s;r)],\nonumber \\ &&
\lambda_v^{qc} (k,s;r)\, = \, \Delta [j +l_0 - (j - v)\cos 2r]+
\nonumber \\ &&+ 2|g| \sum_{f=0}^s \sqrt{(s-f)(f+1)2(2k+1+2f)}
\,d^{j }_{-j+f, -j+v} (2r) \, d^{j}_{-j +f+1, -j +v} (2r)
\nonumber \\ && \approx \Delta [j +l_0 - (j - v)\cos 2r]-
\nonumber \\ &&- 2|g| (j - v) \sin 2r \sqrt{2 [s+2k+1+(- s+2v)\cos
2r]} = \lambda_v^{cmf}(k,s;r),\qquad \label{3.6}\end{eqnarray}
where $d^{j}_{m,n} (2r)$ are  the $SU(2)\, d$- functions
expressed in terms
 of the Gauss hypergeometric function $~_2 F_1$ \cite{11}.
 Approximate values $\lambda_v^{cmf} (k,s;r)$ in (\ref{3.6}) are calculated
 in the cluster mean-field approximation:
 $\langle k,s;f|F (Y_{\alpha})|[k,s;f\rangle=
 F(\langle k,s;f| Y_{\alpha}|k,s;f\rangle)$, and values of the parameter
 $r$ in (\ref{3.5a})-(\ref{3.6}) are found from energy-stationarity-conditions and/or
 from minimizing a proximity measure between exact  Hamiltonian $H$ and its
 $SU(2)$ - quasiclassical approximation
 \begin{equation}H^{qc}(\xi)=\sum_{v,k,s} \, E^{qc}_v(k,s; \xi)\;
 |E^{qc}_v(k,s;\xi)\rangle\;\langle E^{qc}_v (k,s;\xi)|. \label{3.7}\end{equation}
 A standard measure for such estimates on the subspaces $L(k,s)$
 is  defined with the help of the unitarily invariant euclidean operator
 norm \cite{24} as  follows  \cite{11}
 \begin{equation}\delta^2_H (k,s)=\frac{Tr_{(k,s)}(H - H^{qc}(\xi))^2} {Tr_{(k,s)} (H -
 C(l_1))^2}=\frac{\sum_{v}\left[(\lambda_v (k,s))^2
 -(\lambda_v^{qc} (k,s;r))^2\right]}{\sum_{v}(\lambda_v (k,s))^2}.
 \label{3.8}\end{equation}

 This approximation  has been used in \cite{11} for calculating approximate
 expressions $\langle Y_0(t)\rangle^{qc}$ of the temporal dependences
 $\langle Y_0(t) \rangle$ determining, in accordance with Eqs. (\ref{2.1}), (\ref{3.5b}), the dynamics
 of the field-mode  populations:
 \begin{equation} \langle N_0(t) \rangle=\frac{\bar s}{2} - \langle Y_0(t) \rangle, \quad
  \langle N_1(t) \rangle={\bar s} +{\bar k} + 2\langle Y_0(t) \rangle,
  \label{3.9}\end{equation}
%%%%
for different types of initial states; here $\bar s= Tr [\rho
(R_1 -R_0)],\, \bar k= Tr [\rho K ]$ and the quantity $\langle
Y_0(t)\rangle$ is calculated with the help  of Eq. (\ref{2.7}) for
$O=Y_0$. Specifically, in the case of the $SU(2)$-quasiclassical
cluster initial state $| \rangle_C $ of the form (\ref{3.5a}), (\ref{3.5b}) with
$\xi=0, v=0$, belonging to a fixed "optical atom" $L(k,s)$ with
$\langle N_0(0) \rangle = s,  \langle N_1(0) \rangle = k$, Eq.
(\ref{2.7})
 yields an approximate analytical expression \cite{11}
  \begin{eqnarray}
 \langle Y_0(t)\rangle_C^{qc}\approx - \frac{1}{2}\left\{ 1+(s-1)
  {\cal A}_{k,s} (t)\cos [\Omega_L (k,s)t - (s-1)\Phi_{k,s}(t)]\right\}
   \label{3.10a}\end{eqnarray}
 with
 \begin{eqnarray}
  \Omega_{L}(k,s)=
 4|g|\sqrt{\left(1-\frac{1}{s}\right)\left(\frac{s}{2}+k
 +\frac{1}{2}\right)},
 \quad
  \Omega_{l}(k,s)=\frac{4|g|\sqrt{s-1}}{s\sqrt{2s+4k+2}},
  \nonumber \\
 \!\!\!\!\!\!{\cal A}_{k,s} (t)=\left [\cos^2 \Omega_{l}(k,s) t+
 \frac{\sin ^2\Omega_{l}(k,s) t}{s}\right ]^{\frac{s-1}{2}}\!\!\!\!\!\!,
 \quad
 \!\tan \Phi_{k,s} (t)= \frac{\tan\Omega_{l}(k,s) t}{\sqrt
 s},  \label{3.10b}\end{eqnarray}
 which exhibits a high-frequency ($\Omega_{L}(k,s)$) periodic dynamics
 with a slow  ($\Omega_{l}(k,s)$) periodic modulation in the phase and
 amplitude, i.e. an occurrence of a specific temporal coherent structure
  (described in terms of elliptic functions too) \cite{11,12}.
 At the same time for general initial states $| \rangle \in L(H)$ and
 having non-zero projections on all subspaces $L(k,s)$, e.g., for
 Glauber coherent states \cite{11}, analogous calculations lead to series
 containing weighted sums of terms like those given by Eqs. (\ref{3.10a}),(\ref{3.10b}) that
 corresponds to occurrences of coherence-decoherence phenomena like
 "collapse-revivals" revealed in \cite{18,20} by means of other methods.

 However, according to the general quasiclassicality theory \cite{25} all
 approximations (\ref{3.5a})-(\ref{3.6}) (and, hence, (\ref{3.10a}), (\ref{3.10b})) are valid only for large
 values of $s$, and, besides, the measure (\ref{3.8}) gives only a global
 rather than local characteristic of the approximate energy
 spectra $\{E^{qc}_v(k,s;\xi)\}$ that does not allow to feel
 their important symmetry properties and local peculiarities
 related to "energy errors"
 \begin{equation}\Delta E_v(k,s) = \hbar\,
 [\lambda_v(k,s)\,-\,\lambda_v^{qc}(k,s;r)] \equiv \delta
 E_v(k,s) \cdot E_v(k,s). \label{3.11}\end{equation}
 Therefore, we implemented  numerical comparisons of
 both exact and approximate results in order to estimate the
 applicability  range of the quasiclassical approximation (\ref{3.5a})-(\ref{3.6}).

\section{ Comparison of exact and approximate
 calculations in the resonance case}

In order to examine the efficiency of calculation schemes and the
algorithm given above we tested them  by means of computer
experiments for the resonance case  determined by $H$ from Eq.
(\ref{3.1a}) with $\omega_0=2\omega_1,C(l_1)=\omega_1(k+2s), \Delta=0$.

First of all we calculated exact values
$\lambda_f(k,s),\,Q_f^{v}(k,s)$ according to the algorithm of
Section 3 and their approximations $\lambda_v^{cmf}(k,s;r_i),
S^j_{f v}(\xi)$ according to  Eqs. (\ref{3.5a})-(\ref{3.6}) for $ g=1,  k=0,1,
s=20,10^2,5\cdot 10^2,10^3, 10^4$. Values of the fitting
parameter $r$ were determined from
energy-stationarity-conditions: $2r_1= \arccos \frac{1}{3}$
(optimizing only the upper part of spectra) \cite{11}, $2r_3=\arccos
0=\frac{\pi}{2}$ (quasi-linear approximation) \cite{14} and from
minimizing the proximity measure (\ref{3.8}): $2r_2= \arccos
\frac{1}{\sqrt{s}}$ ("smooth" cluster mean-field approximation)
\cite{11}; herewith $\lambda_v^{cmf} (k,s;\mp r_1)$ means that we
take  $r=-r_1$ in the first half of spectra and  $r=r_1$ in the
second one. To estimate the accuracy of approximations we also
used non-invariant measures
\begin{eqnarray}&&\delta^2_E(k,s) =\frac{\sum_{v=0}^s [(\lambda_v (k,s) -
\lambda_v^{cmf}(k,s;r)]^2} {\sum_{v=0}^s (\lambda_v
(k,s))^2},\nonumber\\ && \delta^2_{E_{up}}(k,s)
=\frac{\sum_{v=s/2}^s [(\lambda_v (k,s) -
\lambda_v^{cmf}(k,s;r)]^2} {\sum_{v=s/2}^s (\lambda_v (k,s))^2},
 \label{4.1}\end{eqnarray}
to characterize more precisely (in comparison with Eq. (\ref{3.8}))
energy spectra and standard (related to the Fubini-Study metric
\cite{26} in $L (k,s)$) measures
\begin{equation}\cos ({\bf S},{\bf Q})_{[l_i];v}\equiv \sum_{f}\, S_{fv}^j\, Q_f^v (k,s),
\quad \delta^2_{ef}(k,s;v)= 1-|\cos ({\bf S},{\bf Q})_{k,s;v}|^2
 \label{4.2}\end{equation}
(or associated graphic representations via "overlap areas") to
estimate an "approximation quality" for eigenfunctions. Some of
typical results of these numerical calculations are presented
 in Table 1  and Figs.1,2.\\

{Table 1. Multiplets $\{\lambda_v(k=0,s)\}^s_{v=0},\,
\{\lambda_v^{cmf} (0,s;r_i)\}^s_{v=0}$ with the level step
$\Delta v =10$
 for $s=10^2$.}

\begin{tabbing}
  $v$ \quad\= $\lambda_v(0,s)$\qquad\=$\lambda_v^{cmf}(0,s;r_1)$
\quad\=$\lambda_v^{cmf}(0,s;\mp r_1)$\quad\=$\lambda_v^{cmf}(0,s;
r_2)$ \quad\=$\lambda_v^{cmf}(0,s;r_3)$ \\
  0 \>      -1536.9\>  -1096.7\>  -1545.3\>  -1482.4\>  -1421.2\\
 10 \>      -1151.7\>   -919.6\>  -1205.2\>  -1175.2\>  -1137.0\\
 20 \>       -798.1\>   -720.0\>   -880.0\>   -873.3\>   -852.7\\
 30 \>       -480.3\>   -499.3\>   -570.2\>   -576.7\>   -568.5\\
 40 \>       -205.5\>   -259.0\>   -276.7\>   -285.6\>   -284.2\\
 50 \>          0.0\>      0.0\>      0.0\>      0.0\>      0.0\\
 60 \>        205.5\>    276.7\>    276.7\>    280.0\>    284.2\\
 70 \>        480.3\>    570.2\>    570.2\>    554.3\>    568.5\\
 80 \>        798.1\>    880.0\>    880.0\>    822.8\>    852.7\\
 90 \>       1151.7\>   1205.2\>   1205.2\>   1085.5\>   1137.0\\
100 \>       1536.9\>   1545.3\>   1545.3\>   1342.3\>   1421.2\\
$\delta^2_H=$     \> \>   10.222\>  -12.220\>    0.010\>   -1.000\\
$\delta^2_E=$     \> \>    2.563\>    0.670\>    0.806\>    0.657\\
$\delta^2_{E_{up}}=$\> \>    0.670\>    0.670\>    0.944\>    0.657\\
\end{tabbing}

\begin{center}
\begin{figure}
\epsfig{width=11.1cm,height=6.8cm,file=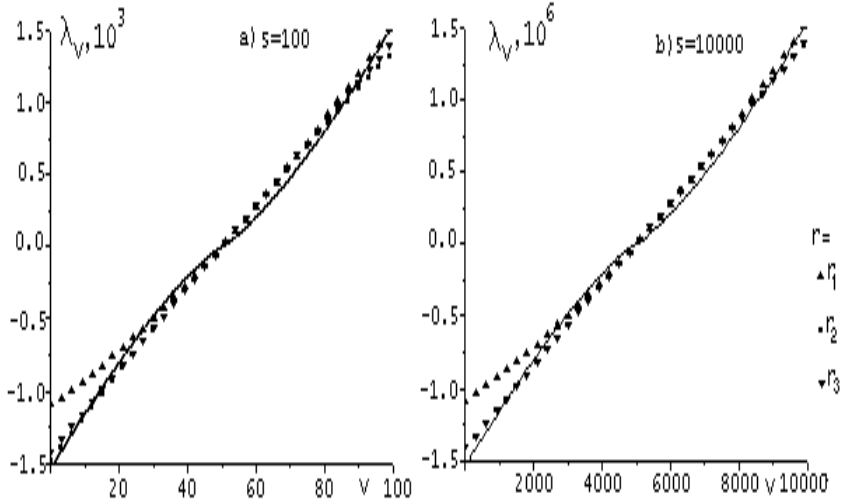} \caption {
Energy levels $\lambda_v(k=0,s)\}$ (solid line) and their
approximations (\ref{3.6}) with $r=r_{i=1,2,3}$ plotted against
the energy label $v$ for $s=10^2$ ( a)\,) and $s=10^4$ ( b)\,).}
\end{figure}
\end{center}

As is seen from data given in Table 1 and Fig. 1 we have an
acceptable consent of exact eigenenergies and their
approximations (\ref{3.6})  at $s \gg 1$ almost everywhere for
$r=r_{i=2,3}$ and $r=\mp r_1$. Discrepancies between exact and
approximate results in the middle parts of spectra, probably, are
due to the availability of the square-root singularities  in the
model Hamiltonian (\ref{2.3a}) re-written (with the help of Eqs. (\ref{3.5b}))
in terms of $Y_{\alpha}$ that is, actually, ignored in the
"smooth" the $su(2)$ - quasiclassical approximation (\ref{3.5a})-(\ref{3.6})
\cite{12}. (Note also that some negative values of $\delta_H^2$ are due
to using $\lambda_v^{cmf}$  instead of $\lambda_v^{qc}$ in Eq.
(\ref{3.8}) and because of calculation errors). However, the
approximation with $r=\mp r_1$ breaks the orthogonality of
eigenfunctions belonging to opposite ends of spectra whereas the
quasi-linear approximation with $r= r_3$ leads to equidistant
spectra within fixed subspaces  $L(k,s)$.
 Therefore, in spite of the
spectrum symmetry breaking, the most satisfactory quasiclassical
approximation is given by Eqs. (\ref{3.5a})-(\ref{3.6}) with $r=r_{2}$ that
minimizes  $\delta_H^2$. Note that the  spectrum asymmetry  at
$r=r_{2}$ and related shifts between amplitude values
$Q_f^{v}(k,s)$ and $S^j_{f v}(\xi)$ (see Fig. 2) are due to using
"smooth" $su(2)$ - quasiclassical eigenfunctions (\ref{3.5a}) \cite{12}.

\begin{center}
\begin{figure}
\epsfig{width=11.1cm,height=6.8cm,file=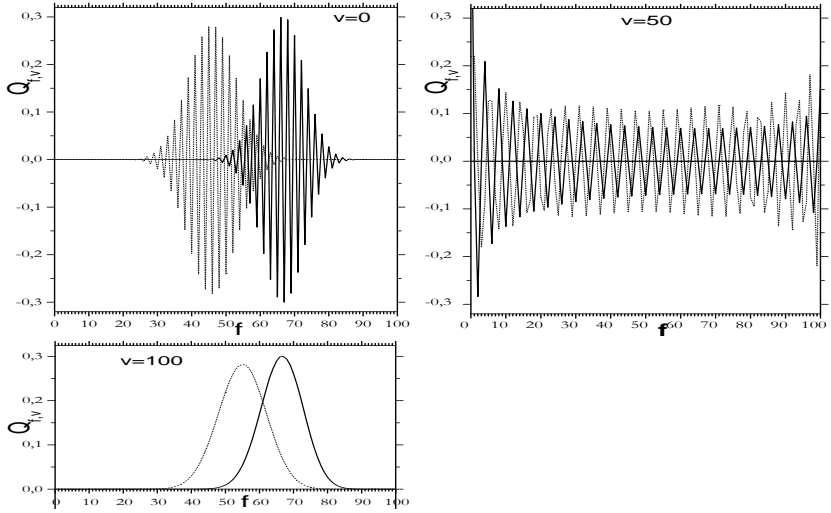} \caption
{Amplitudes $Q^v_f(k,s) $ as functions in $f=0,...,s$ at $k=0,
v=0,50,100$ (solid lines) and
 their quasiclassical approximations $S^j_{vf} (r_2 g/|g|) $ (dashed lines).}
\end{figure}
\end{center}

Besides the verifications above we also performed calculations of
temporal dependences of the quantity $\langle Y_0(t) \rangle_C$
and related dynamics of the normalized average photon numbers
$\langle N_i (t) \rangle_C / s$. Herewith exact dependencies were
calculated with the help of routine package above, whereas
approximate calculations were implemented using approximate
expressions (\ref{3.5a}), (\ref{3.5b}), (\ref{3.6}) for eigenvalues and eigenfunctions.
Results of such calculations for $s=10^2, k=0$ against the
dimensionless time $\tau = g t \sqrt{2 s}$ are plotted in Fig.  3
where we compare exact results with the quasiclassical
approximations obtained with the help of
Eqs. (\ref{3.5a})-(\ref{3.6}) with $r=r_{i=1,2}$ and (\ref{3.10a})-(\ref{3.10b}).\\

\begin{center}
\begin{figure}
\epsfig{width=11.1cm,height=6.8cm,file=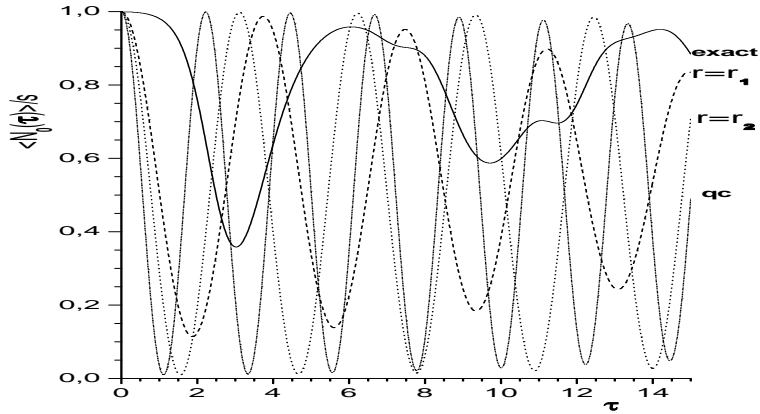} \caption{
 Normalized average photon number $\langle N_0 (t)
\rangle_C / s$ in the fundamental mode  plotted against time
$\tau$
 for $k=0, s=10^2$: (exact) is results of (\ref{3.9})
 and their
 approximations (qc) and ($r=r_{i=1,2}$) calculated
 by  (\ref{3.10a}) and (\ref{3.5a})-(\ref{3.6}) with $r=r_{i=1,2}$.}
\end{figure}
\end{center}

Evidently, the graphic representations of Fig.3 enable us to
reveal transparently a double-periodic component in the exact
multi-frequency dynamics of $\langle Y_0(t)\rangle$ and $\langle
N_0 (t)\rangle$ that is rather well described by Eqs.
(\ref{3.10a}), (\ref{3.10b}) or (\ref{3.5a})-(\ref{3.6}) at
$r=r_{2}$ (and at $r=r_{1}$ ). Note that an availability of this
important dynamic feature is displayed clearer when the
characteristic parameter $s$ increases (in accordance with the
general quasiclassical theory \cite{25}).

\section{ Conclusion}
So, our numerical calculations given in Section 4 show a rather
good qualitative consent of exact and  approximate results at $s
\gg 1$ and at relevant choices of the fitting parameter $r$ in
(\ref{3.5a})-(\ref{3.6}). However, partial quantitative discrepancies of them
require  further improvements of the quasiclassical
approximations used. In particular, the approximate  solutions of
the eigenproblem (\ref{2.6}) can be improved by means of: 1) using less
smooth (in comparison with (\ref{3.5a}), (\ref{3.5b})) generalized coherent states of
the $su_{pd}(2)$ algebra as quasiclassical eigenfunctions (cf.
\cite{12}) and 2) exploiting the standard \cite{24} or special (e.g.,
developed in \cite{27,28}) algebraic perturbative and iterative
algorithms or  modifications of the  algebraic "dressing" schemes
\cite{6}. Then these improvements (along with the exact  calculation
schemes developed above) can be used for a more detail analysis
(like those implemented in \cite{7,8,9,16,17,18,19,20}) of the model under
consideration in all ranges of the parameter $s$ and for
arbitrary initial states. It is also of interest to compare
results obtained (and their improvements) with those of \cite{11}
based on an alternative $SU(2)$ form of $U_H((t)$ as well as
with  calculations performed in \cite{29} using the formalism of the
$q$ - deformed Lie algebra $su_q(2)$. The work along these lines
is in progress.

\end{document}